\documentclass[sigconf, nonacm]{acmart}
\usepackage[a-2b]{pdfx}

\usepackage{tcolorbox}

\setcopyright{none}

\usepackage{multirow}
\usepackage{paralist}

\newcommand{\sstitle}[1]{\noindent\textbf{#1.\/}}

\usepackage{pifont}
\newcommand{\cmark}{\ding{51}} %
\newcommand{\xmark}{\ding{55}} %

 \def\Snospace~{\S{}}

\usepackage{xspace}
\newcommand{\toolname}{\texttt{GRAST-SQL}\xspace}

\begin{document}

\setlength{\belowdisplayskip}{3pt}
\setlength{\belowdisplayshortskip}{3pt}
\setlength{\abovedisplayskip}{3pt}
\setlength{\abovedisplayshortskip}{3pt}

\title{Scaling Text2SQL via LLM-efficient Schema Filtering with Functional Dependency Graph Rerankers}

\author[Thanh Dat Hoang et al.]{Thanh Dat Hoang$^1$, Thanh Tam Nguyen$^1$,
Thanh Trung Huynh$^2$, \\ Hongzhi Yin$^3$, Quoc Viet Hung Nguyen$^1$}
\affiliation{
\institution{$^1$Griffith University (Australia), $^2$VinUniversity (Vietnam),
$^3$The University of Queensland (Australia)}
\country{\vspace{1em}}
}

\begin{abstract}
Most modern Text2SQL systems prompt large language models (LLMs) with entire schemas -- mostly column information -- alongside the user's question. While effective on small databases, this approach fails on real-world schemas that exceed LLM context limits, even for commercial models. The recent Spider 2.0 benchmark exemplifies this with hundreds of tables and tens of thousands of columns, where existing systems often break. Current mitigations either rely on costly multi-step prompting pipelines or filter columns by ranking them against user's question independently, ignoring inter-column structure. To scale existing systems, we introduce \toolname, an open-source, LLM-efficient schema filtering framework that compacts Text2SQL prompts by (i) ranking columns with a query-aware LLM encoder enriched with values and metadata, (ii) reranking inter-connected columns via a lightweight graph transformer over functional dependencies, and (iii) selecting a connectivity-preserving sub-schema with a Steiner-tree heuristic. Experiments on real datasets show that \toolname achieves near-perfect recall and higher precision than CodeS, SchemaExP, Qwen rerankers, and embedding retrievers, while maintaining sub-second median latency and scaling to schemas with 23,000+ columns. Our source code is available at \url{https://github.com/thanhdath/grast-sql}.

\end{abstract}

\maketitle

%

\section{Introduction}

\begin{table*}[t]
\centering
\small
\caption{Functional comparison. Max cols$^\ast$ reports raw capacity without any heuristics including chunking, values are approximate and depend on schema tokenization (e.g., name lengths); Scope: Single/Multi table; Conn. refers to Connectivity; Relation modeling denotes how column relations are learned. In our experiments, \toolname scales to 23{,}067 columns on Spider 2.0-lite.}

\label{tab:sf-compact-scope}
\vspace{-.2cm}
\setlength{\tabcolsep}{4pt}
\begin{tabular}{l c c c c c c l l c}
\hline
Method & Scope & Names & Meaning & Values & Types & Conn. & LM (architecture) & Relation modeling & Max cols$^\ast$ (no heuristics) \\
\hline
\textbf{\toolname (ours)} & \textbf{Multi} & \cmark & \cmark (long) & \cmark & \cmark & \textbf{\cmark} & Qwen3 (decoder) & Graph & \textbf{23{,}067} \\
SchemaExP & Single & \cmark & \xmark & \cmark & \cmark & \xmark & BERT (encoder) &  LM attention& $\approx$ 90 \\
RESDSQL & Multi & \cmark & \xmark & \xmark & \xmark & \xmark & RoBERTa-L (encoder) &  LM attention & $\approx$ 159 \\
CodeS & Multi & \cmark & \cmark (short) & \xmark & \xmark & \xmark & RoBERTa-L/XXL (encoder) &  LM attention & $\approx$ 117 \\
PURple-SQL & Multi & \cmark & \cmark (short) & \xmark & \xmark & \cmark & RoBERTa-L (encoder) &  LM attention & $\approx$ 117 \\
\hline
\end{tabular}
\vspace{-.2cm}
\end{table*}

Text2SQL, the task of translating natural-language questions into executable SQL queries, is a long-standing challenge in database applications \cite{zhong2017seq2sql,xu2017sqlnet,yu2018typesql,yu2018syntaxsqlnet}. 
While recent advances in large language models (LLMs) have brought notable improvements \cite{pourreza2023dinsql,sun2023sqlpalm,dong2023c3,liu2023comprehensive}, the task remains difficult due to the increasing complexity of user queries and the scale of modern database schemas \cite{rokis2022challenges}.
Most current approaches prompt LLMs with extensive schema information -- including table and column names, relations, and even values -- alongside the user's question to generate the SQL query \cite{roziere2023code,li2023starcoder}. 

However, recent benchmarks such as Spider~2.0~\cite{lei2024spider} include databases with hundreds of tables, tens of thousands of columns, and extensive metadata, exposing their scalability issues. First, including all schema elements inflates prompt length, often pushing or exceeding LLM context limits and significantly increasing token costs~\cite{lei2024spider}. Second, it adds noise, making the model more likely to focus on irrelevant tables or columns, which can lead to incorrect join paths or attribute selections in the generated SQL~\cite{li2023resdsql, lei2024spider}.

\begin{figure}[t]
    \centering
    \includegraphics[width=1.0\linewidth]{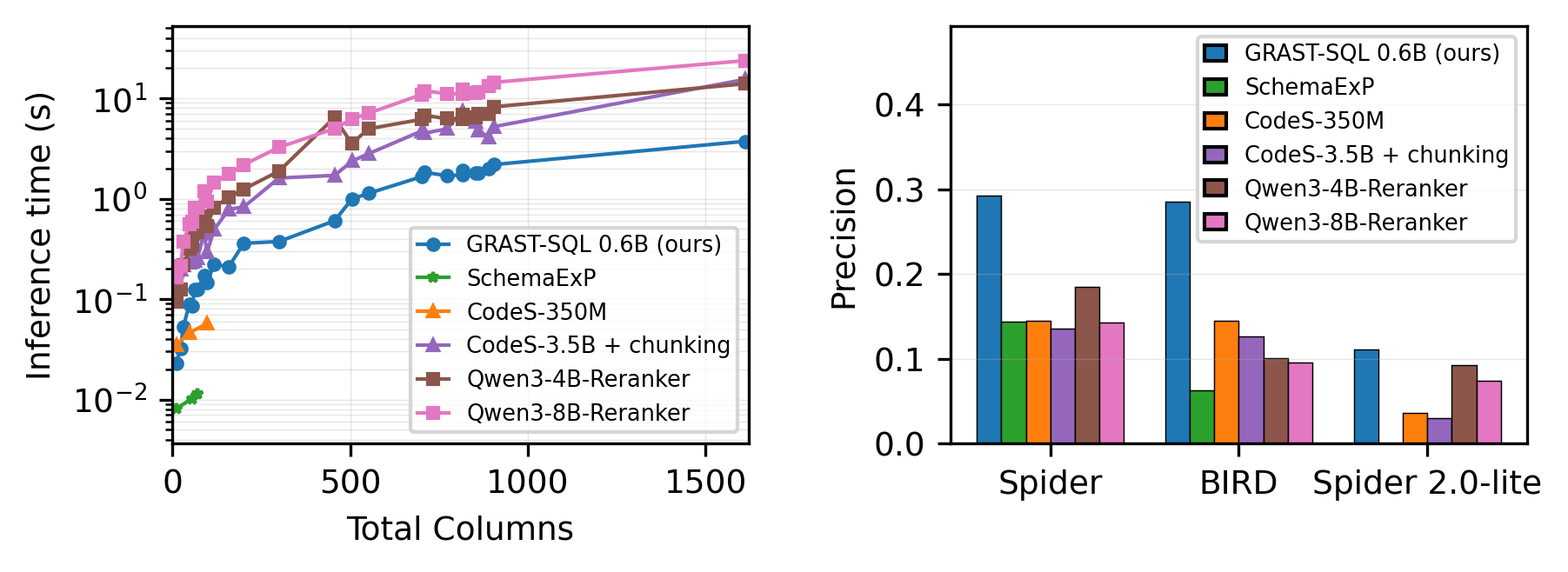}
    \caption{Scaling but better.
    \emph{Left:} Inference time vs. total columns on BIRD and Spider 2.0-lite. Prior rankers (e.g., SchemaExP, CodeS w/o chunking) fail beyond $\sim$90-160 columns, while \toolname scales smoothly.
\emph{Right:} Precision on Spider, BIRD, and Spider 2.0-lite, where \toolname consistently outperforms by exploiting column semantics.}
    \label{fig:motivating}
\end{figure}

Existing scaling mitigations address these issues in two main ways. The first one relies on commercial models with large context windows, which require costly multi-step prompting pipelines since a single flat prompt can still exceed their limits, making them expensive to deploy~\cite{pourreza2023dinsql,pourreza2025chasesql,talaei2024chess,lei2024spider}. For instance, CHESS-style schema filtering on the BIRD dev set consumes $\sim$340K tokens/request ($\approx\$3.4$ with GPT-4-turbo), let alone when dealing with 1,000 requests \cite{chung2025long}. On complex databases like Spider~2.0, some databases can even exceed 20M tokens and require strong reasoning models such as GPT-o3~\cite{deng2025reforce}. The second way, a more affordable alternative, uses open-source schema filtering with relation-aware ranking (RESDSQL, CodeS, SchemaExP, DCG-SQL, PURple), but still limited by context length and thus cannot fully exploit rich column semantics such as metadata and representative values~\cite{li2023resdsql,li2024codes,lee2025dcg,ren2024purple}. Additionally, one might view column filtering like document retrieval in RAG pipelines, but naive RAG-style embedding and rerankers ignore dependencies across columns, producing subsets that appear relevant but omit essential fields, making SQL generation infeasible. These issues are amplified on Spider 2.0 and BIRD, where wide schemas often drop key columns, retain noise, and break joinability.

To address these limits, we present \toolname, a compact, open-source pipeline for LLM-efficient schema filtering that avoids oversized commercial models while scaling to massive, real-world databases. \toolname first enriches each database with a functional dependency (FD) graph and concise metadata. Given a question, it (1) initializes query-aware column embeddings with an instruction-tuned LLM, (2) refines them via a relation-aware graph transformer over the FD graph, and (3) applies a Steiner-tree spanner to guarantee valid joins under a top-\(K\) cut. The result is a small sub-schema that preserves essential join structure and semantics while reducing token cost and noise.
As in \autoref{fig:motivating}(left), existing schema filters (e.g., CodeS, SchemaExP) unable to handle wide schemas: context windows are exceeded once databases contain a few hundred columns. In contrast, see \autoref{fig:motivating}(right), our approach preserves smooth latency as schema size grows and achieves higher precision by capturing column semantics rather than relying only on names.

Our main contributions are summarized as follows.
\begin{compactitem}
\item We present \toolname, a lightweight, LLM-efficient schema filtering framework that unifies query-column ranking, graph-based reranking, and connectivity enforcement, designed to operate with compact models.

\item We introduce a functional dependency graph that encodes primary-key, foreign-key, and intra-table relations, together with a graph-based reranker that learns and propagates column relationships across this structure, producing more accurate column scoring while ensuring valid join paths.

\item We demonstrate that \toolname outperforms existing baselines, scaling to enterprise-scale databases with hundreds of tables and thousands of columns, achieving sub-second median latency with near-perfect recall, higher precision, and superior ROC/PR AUC, while reducing prompt length for Text2SQL by up to 50\% on BIRD.

\item We release training and evaluation datasets for schema filtering on Spider, BIRD, and Spider 2.0-lite, to enable further research and comparison.

\item We open-source the trained models together with the full codebase and scripts, supporting reuse on custom databases or re-training on new datasets.

\end{compactitem}
In the following, we review related work (\autoref{sec:related}), present our methodology (\autoref{sec:method}), report experimental results (\autoref{sec:experiments}), and conclude (\autoref{sec:conclusion}).

\section{Related Works}
\label{sec:related}

\sstitle{Text2SQL}
Modern Text2SQL began with neural parsers like SyntaxSQLNet \cite{yu2018syntaxsqlnet} and RAT-SQL \cite{wang2020rat}, which encoded full schemas, followed by constrained decoding with PICARD \cite{scholak2021picard}. With growing schema sizes, filtering became essential: RESDSQL and CodeS rank elements by question relevance \cite{li2023resdsql, li2024codes}, while PURple-SQL ensures coverage via connectivity-aware pruning. More recent systems such as CHESS \cite{talaei2024chess}, CHASE-SQL \cite{pourreza2025chasesql}, and Alpha-SQL \cite{li2025alphasql} use prompting, planning, and reranking in multi-stage pipelines. Across these methods, schema filtering remains central to reducing noise and token costs when using proprietary LLMs \cite{li2024codes,lee2025dcg}, though many solutions rely on costly long-context models---raising concerns over latency, cost, and data privacy---or simple heuristics that miss critical columns, thereby reducing execution accuracy on complex schemas.

Our work goes beyond the state-of-the-art by proposing an open-source lightweight ranker that reduces unnecessary token costs while preserving Text2SQL accuracy on large databases.

\sstitle{Schema Filtering}
Schema filtering selects only tables and columns relevant to a question, cutting prompt length and noise while preserving joinability. Early methods relied on schema/value linking and structure-aware encoders: IRNet links mentions to schema elements \cite{guo2019towards}; graph-based encoders model database structure \cite{bogin2019representing,bogin2019global}; RAT-SQL applies relation-aware attention \cite{wang2020rat}; BRIDGE augments fields with values \cite{lin2020bridging}; and ShadowGNN improves linking under noisy forms \cite{chen2021shadowgnn}. Later, RESDSQL and CodeS rank schema elements with transformers \cite{li2023resdsql,li2024codes}, DCG-SQL prunes and refines via a schema graph \cite{lee2025dcg}, and PURple enforces connectivity with a Steiner-tree objective \cite{ren2024purple}. Recent agent frameworks such as CHESS and MAC-SQL integrate schema selectors but rely on repeated calls to large proprietary LLMs \cite{talaei2024chess,wang2025mac}. Overall, schema filtering must capture column relationships -- ranking independently is insufficient -- yet lightweight methods remain limited by context length, often discarding key fields or retaining excess noise without guarantees of coverage or calibrated scoring.

Overcoming these limitations, we propose an LLM-efficient filter that handles long-context column semantics while capturing column relationships, and scales to thousands of total columns. (See our functional comparison in \autoref{tab:sf-compact-scope}).

\section{Methodology}
\label{sec:method}

\begin{figure*}[t]
  \centering
  \includegraphics[width=1\linewidth]{main-flow.png}
  \vspace{-.6cm}
\caption{Overview of \toolname. 
Inputs: database and metadata. 
Schema Enricher: constructs FD graph and enriches Metadata. 
Query-aware Column Encoder: inits column embeddings via instruction-tuned LLM. 
Graph-based Reranker: enhances column embeddings via relationships in the FD graph. 
Steiner Tree Spanner: ensures valid joins under top-K selection.
}
  \label{fig:framework}
  \vspace{-.3cm}
\end{figure*}

\subsection{Problem Formulation}
\label{sec:problem_formulation}
The schema filtering problem in Text2SQL focuses on identifying the minimal subset of a database schema necessary to accurately translate a natural-language query into an executable SQL statement. Let \(D = (T, C, F)\) represent a relational database, where \(T\) is the set of tables, \(C\) the set of columns, and \(F\) the set of foreign key relations. Given a query \(q\), the objective is to generate a SQL query \(s\) whose execution returns the correct result. The role of schema filtering is to select a sub-schema \((T^{\star}, C^{\star}, F^{\star})\) that includes only the elements relevant to answering \(q\). This serves two purposes: reducing noise that misleads SQL generation and avoiding unrelated components that inflate LLM token costs. An effective schema filter must balance coverage and conciseness, preserving essential schema components while suppressing irrelevant ones.

\begin{example}
  \label{ex:example1}
Consider a database $D$ with tables  
Students, Enrollments, Courses, 
Departments, and others. For a question $q$: 
``Count the number of courses offered in the 
\emph{Computer Science} department,'' the sufficient  
sub-schema is
$T^\star=$ \{Courses, Departments\},
$C^\star=$ \{Courses.cid, Courses.dept\_id, Departments.did, Departments.name\}, 
and the foreign-key set is 
$F^\star=$ \{Courses.dept\_id $\to$ Departments.did\}. Other tables are irrelevant and can be pruned, reducing noise and token cost while still preserving the correct join path and full query semantics. The corresponding SQL is:
\begin{verbatim}
SELECT COUNT(c.cid) AS num_courses
FROM Courses AS c
JOIN Departments AS d
  ON c.dept_id = d.did
WHERE d.name = 'Computer Science';
\end{verbatim}
\end{example}

\noindent
\autoref{fig:framework} presents our four-stage approach, \toolname, to schema filtering. Each stage is detailed in the following sections.
Our methodology goes beyond heuristics or long-context LLMs by combining query-aware ranking, functional dependency graph reasoning, and Steiner-tree connectivity. Its implementation is challenging, requiring key recovery in incomplete schemas, graph-based reranking, Steiner-tree adaptation, and engineering for scale.

\subsection{Schema Enricher}
\label{sec:stage1}

Given a SQL database $D$, the Schema Enricher builds (1) an FD graph over columns and (2) an enriched metadata map, both constructed once per database for downstream reuse. 

\subsubsection{Functional Dependency Graph Construction}
\label{sec:fd-graph}

We construct a column-level graph \(G = (V, E)\), where each node \(v \in V\) corresponds to a column \(c \in C\), and each edge \(e \in E\) encodes a schema dependency, either declared in the database or predicted. The purpose of this graph is to ensure structurally important columns are not discarded during filtering, even when they have no lexical or semantic overlap with the natural-language query. As illustrated in \autoref{ex:example1}, key columns such as \texttt{cid} and \texttt{did} are indispensable for joins and aggregations, yet bear no surface-form relation to the query text; retaining them requires structural connectivity rather than purely semantic retrieval.

\sstitle{Edge types}
The graph includes three types of dependencies:
\begin{compactitem}
\item \emph{Foreign key edges}: connect a source column to its referenced column via a foreign key relation, as defined in \(F\).
\item \emph{Column-to-foreign-key edges}: connect every non-key column in a table to the corresponding foreign key column of that table or the referenced table.
\item \emph{Column-to-primary-key edges}: connect non-primary-key columns in a table to its table’s primary key.
\end{compactitem}

\sstitle{Predicting missing keys}
In practice, many real-world databases (notably those in BIRD and Spider~2.0) omit explicit primary or foreign key declarations. To recover this missing structure, we use OpenAI prompting (GPT-4.1-mini) with schema metadata as input. These predictions are used solely for enriching the graph and do not modify the underlying database. This is feasible because many keys follow consistent naming conventions, such as primary keys ending with \texttt{id}, and relationships can often be inferred from reused identifiers across tables, e.g., locations.id $\to$ cards.location\_id.

\begin{compactitem}
\item \textit{Primary keys:} The prompt receives a list of tables along with their column names, types, and meanings. The task is to identify a list of primary keys, and return the output in JSON format. The predicted primary keys are then merged with any declared keys to produce \(K_t\) for each table.
    
\item \textit{Foreign keys:} Using the same schema listing, the prompt asks the model to return a JSON list of likely foreign-key relationships in the form \((t_u.c_u \rightarrow t_v.c_v)\). Predicted links are merged with the declared foreign keys \(F\) to form \(\widehat{F}\), which is only used to enrich the connectivity of the graph.
\end{compactitem}

\vspace{-0.8em}
\subsubsection{Metadata Enrichment}
\label{sec:metadata-completion}

While structural connectivity ensures valid join paths and prevents the pruning of critical key columns (e.g., primary and foreign keys), semantic understanding of tables and columns remains essential for identifying which columns are relevant to a natural-language query. To support this, we enrich each column with concise, human-readable descriptions.

\vspace{0.5em}
\noindent\textbf{Observed metadata.}
All datasets used in this work -- Spider, BIRD, and Spider~2.0 -- include external metadata files that annotate tables and columns with textual descriptions. Where these descriptions are available, we incorporate them directly without modification.

\vspace{0.5em}
\noindent\textbf{Missing metadata.}
In real-world databases, particularly those in BIRD and Spider~2.0, many table and column descriptions are missing or abbreviated (e.g., \texttt{crs\_cd}, \texttt{std\_nm}), making it difficult for language models to correctly align schema elements with natural-language questions. Providing explicit meanings improves interpretability and retrieval quality. While this task is ideally performed by humans, we automate it in this work using OpenAI prompting (GPT-4.1-mini). We perform metadata enrichment in two cases:

\begin{compactitem}
\item \textit{Table meaning generation:}  
For tables without descriptions, we prompt the model with the database and table name plus a list of columns (name, type, existing description if any, and sample values), asking it to generate a one-sentence summary of the table's purpose or content.

\item \textit{Column meaning generation:}  
For columns lacking descriptions, the input includes the database name, table name, column name, type, and sample values. The model generates a single-sentence explanation of the column’s meaning.
\end{compactitem}

We apply this procedure only to BIRD and Spider 2.0, where missing or ambiguous schema elements are common. This process consumes a total of 22K tokens on BIRD dev (11 databases, $\approx\$0.009$ total) and 32.6M on Spider~2.0-lite (99 databases, $\approx\$13$ total).

\subsection{Question-Aware Column Encoder}
\label{sec:stage2}

Given a query \(q\) and schema \(D\), we build a column context for each \(c\in C\) and convert it into a query–conditioned embedding used to initialise the node feature in the FD graph (see \autoref{fig:framework}). Each column context includes the following components:
\begin{compactitem}
\item Table name: the table \(t \in T\) to which column \(c\) belongs.
    \item Column name: the raw name or identifier of the column \(c\).
    \item Table description: a textual explanation describing the semantics of table \(t\).
    \item Column description: a textual explanation describing the semantics of column \(c\).
    \item Data type: the SQL data type of \(c\), such as \texttt{integer}, \texttt{text}.
    \item Sample values: cell values retrieved by a value retriever selecting entries aligned with the query \(q\).
    \item Missingness flag: a Boolean flag indicating whether \(c\) may contain missing or null entries.
    \item Value description: a textual description of the meaning or encoding of the values stored in \(c\).
\end{compactitem}

\sstitle{Value retriever}
All distinct cell values are pre-indexed offline in a sparse inverted file. At run time, each value \(v\) is scored against the query with BM25, and the top-\(k\) positives (\(k=2\)) are placed in the \emph{Sample values} field, as they provide more useful examples when closely matching the query. If none are found, a representative example is used; for Spider~2.0, we skip indexing and directly use the dataset’s provided sample values.

\sstitle{Embedding extraction}
To obtain a query-aware embedding for a column \(c\), we format the query \(q\) and the context of \(c\) into an instruction-following prompt \(\mathcal{I}(q,c)\). An LLM is asked whether \(c\) is needed to answer \(q\) (see Prompt Template~1, \autoref{lst:prompt_template}). Rather than using the predicted label, we extract the hidden state the model uses to score the first answer token (“yes”/“no”), which reflects its contextual assessment of the relevance of \(c\). The input sequence is:

\[
x_{q,c}
= \langle\texttt{system}\rangle\,\mathcal{I}_{\text{sys}}\,\langle\texttt{end}\rangle\;
  \langle\texttt{user}\rangle\,\mathcal{I}(q,c)\,\langle\texttt{end}\rangle\;
  \langle\texttt{assistant}\rangle .
\]
where $\mathcal{I}_{\text{sys}}$ is the system prompt. The query-aware initial embedding \(h^{(0)}_{v_c}\) for the graph node \(v_c\) is taken as the hidden state at the final input position, the final hidden state before decoding the next token “yes” / “no”. This state, which captures the interaction between \(q\) and column context \(c\), initializes the column node feature and is subsequently refined by the graph-based module.

\begin{tcolorbox}[width=0.99\linewidth,title=Prompt Template 1: Column–query relevance]
\label{lst:prompt_template}
\footnotesize
\vspace{-0.2cm}
\begin{verbatim}
<|im_start|>system
Judge whether the column (Document) is necessary to use 
when writing the SQL query, based on the provided Query 
and Instruct. The answer can only be "yes" or "no".
<|im_end|>
<|im_start|>user
<Instruct>: Given a natural-language database question 
(Query) and a column description (Document), decide if 
the column may be necessary to answer the question.
<Query>: {q}
<Document>: {context(c)}
<|im_end|>
<|im_start|>assistant
<think>\n\n</think>\n\n
\end{verbatim}
\vspace{-0.2cm}
\end{tcolorbox}

\subsection{Graph-Based Reranker}
\label{sec:graph-refinement}

After constructing the functional dependency graph \(G=(V,E)\) in Stage \autoref{sec:fd-graph}, each column node \(v \in V\) is initialised with the query-conditioned embedding \(h^{(0)}_{v}\) from \autoref{sec:stage2}. To inject structural context, we apply a relation-aware graph transformer~\cite{yun2019graph} to the functional dependency graph. Let the typed-edge set be \(\tilde{E} \subseteq V \times V \times \mathcal{R}\), where
$
\mathcal{R}=\{\text{foreign\_key},\ \text{column}\!\to\!\text{foreign\_key},\ \text{column}\!\to\!\text{primary\_key}\}.
$
Across \(\ell\) layers, the representation of each node \(v\) is updated by attending to its neighbours:
\begin{equation}
h^{(\ell)}_{v}
=
W_{\text{self}}^{(\ell)}\,h^{(\ell-1)}_{v}
\;+\!
\sum_{(u,v,r)\in E}
\alpha^{(\ell)}_{r}(u,v)\,
W_{r}^{(\ell)}\,h^{(\ell-1)}_{u},
\end{equation}
where \(W_{\text{self}}^{(\ell)}\) and \(W_{r}^{(\ell)}\) are trainable weights, and 
\(\alpha^{(\ell)}_{r}(u,v)\) is a relation-specific (scaled dot-product) attention coefficient. 
Edges are treated as directed; in practice we add reverse relations to enable bidirectional propagation.

After the final layer, the refined embedding \(h^{(L)}_{v}\) captures both the column’s query relevance 
and its structural role in the functional dependency graph. 
A linear projection maps this embedding to a relevance score $r(v,q)$ for the column \(c\) corresponding to node \(v\):
\begin{equation}
r(v,q) = w^{\top} h^{(L)}_{v} + b,
\end{equation}
where \(w\) and \(b\) are learnable parameters. 
Training uses a margin-based contrastive loss to rank columns used in the ground-truth SQL above irrelevant ones. 
The resulting scores are passed to Stage~4 to select a connected subset of columns that forms a valid join path.

\subsection{Steiner-Tree Spanner}
\label{sec:steiner}

Although the top-ranked columns from Stage~3 often already include the foreign keys needed for valid join paths, connectivity is not guaranteed under a strict top-\(K\) cut; this stage is an optional add-on that explicitly preserves valid join paths by inserting any missing key columns via a Steiner-tree procedure on the functional dependency graph~\citep{hwang1992steiner, ren2024purple}.

Let \(R=\{v_{1},\dots,v_{m}\}\subseteq V\) be the \(m\) columns with the largest relevance scores \(r(v,q)\). The objective is to find a minimal connected subgraph \(G^{\star}=(V^{\star},E^{\star})\) that spans all terminals \(R\) while keeping the total edge cost small:
\begin{equation}
\min_{V^{\star},E^{\star}\subseteq G}\;\;
\sum_{e\in E^{\star}} w_{e}
\quad\text{s.t.}\quad
R\subseteq V^{\star},\;G^{\star}\text{ is connected}.
\end{equation}

Edge costs are set to \(w_{e}=0\) when \(e\) links a terminal to a primary- or foreign-key column in the same table, and \(w_{e}=1\) otherwise, which biases the solution toward economical joins. We use a greedy approximation that adds, at each step, the cheapest edge that reduces the number of connected components; this is sufficient in practice and runs in near-linear time with respect to \(|E|\).

The auxiliary columns introduced by this procedure are collected in \(C_{\text{aux}}=V^{\star}\setminus R\), and the final filtered schema is \(C^{\star}=R\cup C_{\text{aux}}\). Because the Steiner tree is built over the FD graph, every join key required for a valid SQL query appears in \(C^{\star}\), yielding a concise yet fully connected prompt for the downstream decoder.

\section{Model Training}
\label{sec:training}

\subsection{Training Data Creation}

We construct two training datasets using the train splits of the Spider and BIRD benchmarks. In the original dataset, each example consists of a natural-language question \(q\) and its corresponding SQL query \(y\). To derive column-level supervision:

\begin{compactitem}
    \item We extract the set of columns \(\mathcal{C}^{+}(q)\) that appear in the gold SQL query \(y\).
    \item All other columns in the same database schema are treated as negative candidates, denoted \(\mathcal{C}^{-}(q)\).
\end{compactitem}

Since many questions admit multiple correct SQL queries beyond the labeled one, we augment supervision by generating alternative queries with LLMs (e.g., CodeS \cite{li2024codes}, CHASE-SQL \cite{pourreza2025chasesql}) using temperature 1.0 to return 20 SQL candidates, then select correct execution ones. The union of their columns enlarges the true set \(\mathcal{C}^{+}(q)\). Each training instance is represented as \((q, \mathcal{C}^{+}(q), \mathcal{C}^{-}(q))\), and after filtering misaligned labels, the final training sets contain 9,369 examples from BIRD and 8,395 from Spider.

\subsection{Two-Step Training Procedure}

\sstitle{Step 1: LLM-Reranker Training}
We fine-tune a decoder-only LLM reranker that computes a relevance score \(f(q,c)\) for each column \(c\) conditioned on the question \(q\). The input concatenates the column context (See \autoref{sec:stage1}) and \(q\) in an instruction-style prompt that restricts the answer to “yes” or “no”; \(f(q,c)\) is defined as the unnormalized logit of token \texttt{``yes''} at the first decoding step. Concretely, we instantiate the reranker with \emph{Qwen3-Reranker} \cite{qwen3embedding} in sizes (0.6B, 4B, 8B) and apply LoRA. We allow long contexts by setting the maximum query and passage length to 4096 tokens.

The training objective is a group-wise InfoNCE loss:
\begin{equation}
\mathcal{L}(q)
= - \log \frac{\exp\!\big(f(q,c^{+})\big)}
{\sum_{c \in \{c^{+}\} \cup \mathcal{N}(q)} \exp\!\big(f(q,c)\big)} ,
\end{equation}
where \(c^{+}\in\mathcal{C}^{+}(q)\) is a positive column from the gold SQL \(y\), and \(\mathcal{N}(q)\) are sampled negatives. Each training instance is paired with seven negatives in addition to the positive column, and we further exploit in-batch negatives to increase diversity. Training is conducted for two epochs with an effective batch size of 32 and a learning rate of \(2\times10^{-4}\).

\sstitle{Step 2: Graph Transformer Training}
After training the LLM-reranker, we freeze its parameters and use it to compute initial query-aware embeddings \(h^{(0)}_v\) for each column \(c \in C\), as outlined in \autoref{sec:stage2}. These embeddings are used as node features in the functional dependency graph \(G = (V, E)\). We then train a multi-layer graph transformer as described in \autoref{sec:graph-refinement}, optimizing the following margin-based contrastive loss:
\begin{equation}
\mathcal{L}(q) = \sum_{c^- \in \mathcal{N}(q)} \max \left( 0, \gamma - f'(q, c^+) + f'(q, c^-) \right),
\end{equation}
where \(f'(q, c)\) denotes the final score output by the graph-based model for column \(c\).  
Training is performed for 40 epochs with a batch size of 32 and a learning rate of \(5 \times 10^{-5}\).

\section{Experiments}
\label{sec:experiments}

In this section, we aim to answer the following research questions:
\begin{itemize}
\item[(RQ1)] Does \toolname outperform existing schema filtering baselines in precision and ranking quality? (\autoref{sec:mainresults})
\item[(RQ2)] What benefits do the functional dependency graph and Steiner closure provide? (\autoref{sec:ablation})
\item[(RQ3)] How does \toolname scale to very wide schemas with hundreds of tables and thousands of columns? (\autoref{sec:efficiency})
\item[(RQ4)] How sensitive is it to hyperparameter settings such as layer depth and hidden size? (\autoref{sec:param-sensitivity})
\item[(RQ5)] How robust is \toolname to thresholds? (\autoref{sec:threshold})
\item[(RQ6)] Can \toolname reduce token costs in end-to-end Text2SQL without hurting execution accuracy? (\autoref{sec:end2end})

\end{itemize}

\subsection{Experimental Setup}

\begin{table}[t]
\centering
\small
\caption{Schema width statistics across Spider, BIRD, and Spider 2.0-lite. Medians, 95th percentiles, and maxima are computed per dataset.}
\label{tab:schema-stats-4sets}
\vspace{-.3cm}
\begin{tabular}{lrrrr}
\toprule
Dataset & \shortstack{\#DBs} & \shortstack{Tables\\(med/95p/max)} & \shortstack{Columns\\(med/95p/max)} \\
\midrule
Spider (train) & 146 & 4 / 14 / 26 & 20 / 66 / 352 \\
Spider (dev) & 20 & 3 / 8 / 11 & 18 / 49 / 56  \\
BIRD (train) & 69 & 5 / 19 / 66 & 34 / 126 / 457  \\
BIRD (dev) & 11 & 8 / 12 / 15 & 64 / 159 / 201 \\
Spider 2.0-lite & 99 & 12 / 133 / 381 & 228 / 3910 / 23067  \\
\bottomrule
\end{tabular}
\vspace{-.5cm}
\end{table}

\begin{table*}[!ht]
\centering
\small
\caption{Schema filtering performance comparison on Spider Dev, BIRD Dev, and Spider 2.0-lite public set.}
\label{tab:main-results}
\vspace{-.3cm}
\scalebox{0.99}{
\begin{tabular}{l|cccc|cccc|cccc}
\toprule
\multirow{2}{*}{\textbf{Method}} & 
\multicolumn{4}{c|}{\textbf{Spider Dev}} & 
\multicolumn{4}{c|}{\textbf{BIRD Dev}} & 
\multicolumn{4}{c}{\textbf{Spider 2.0-lite}} \\
\cline{2-13}
& ROC & PR & Recall & Precision 
& ROC & PR & Recall & Precision 
& ROC & PR & Recall & Precision \\
\midrule
\multicolumn{13}{l}{Schema filtering baselines} \\ \hline
SchemaExP & 0.945 & 0.778 & 0.994 & 0.144 & 0.334 & 0.043 & 0.992 & 0.063 & - & - & - & - \\
CodeS-350M & 0.977 & 0.831 & 0.998 & 0.145 & 0.952 & 0.696 & 0.992 & 0.145 & - & - & - & - \\
CodeS-350M + chunking & 0.983 & 0.847 & 0.998 & 0.151 & 0.960 & 0.725 & 0.991 & 0.146 & 0.898 & 0.138 & 0.978 & 0.040 \\
CodeS-3.5B + chunking & 0.983 & 0.868 & 0.997 & 0.136 & 0.972 & 0.755 & 0.985 & 0.126 & 0.912 & 0.143 & 0.901 & 0.030 \\
bge-m3 & 0.836 & 0.427 & 0.959 & 0.145 & 0.799 & 0.346 & 0.991 & 0.095 & 0.744 & 0.041 & 0.991 & 0.078 \\
Qwen3-Embedding-0.6B & 0.841 & 0.453 & 0.998 & 0.159 & 0.70 & 0.148 & 0.991 & 0.109 & 0.681 & 0.035 & 0.991 & 0.081 \\
bge-reranker-v2-m3 & 0.808 & 0.418 & 0.959 & 0.142 & 0.650 & 0.084 & 0.991 & 0.096 & 0.728 & 0.020 & 0.991 & 0.080 \\
bge-reranker-v2-minicpm-layerwise & 0.850 & 0.499 & 0.959 & 0.142 & 0.647 & 0.095 & 0.991 & 0.096 & 0.784 & 0.068 & 0.991 & 0.083 \\
Qwen3-Reranker-0.6B & 0.922 & 0.663 & 0.959 & 0.147 & 0.884 & 0.470 & 0.992 & 0.096 & 0.841 & 0.073 & 0.991 & 0.077 \\
Qwen3-Reranker-4B & 0.944 & 0.723 & 0.959 & 0.185 & 0.939 & 0.612 & 0.992 & 0.101 & 0.915 & 0.214 & 0.991 & 0.093 \\
Qwen3-Reranker-8B & 0.943 & 0.744 & 0.959 & 0.143 & 0.923 & 0.607 & 0.992 & 0.096 & 0.899 & 0.241 & 0.991 & 0.074 \\
\hline
\multicolumn{13}{l}{Ours} \\ \hline
\toolname 0.6B & 0.981 & 0.899 & \textbf{0.998} & 0.293 & 0.978 & 0.777 & \textbf{0.992} & 0.285 & 0.937 & 0.236 & \textbf{0.991} & 0.111 \\
\toolname 4B & 0.987 & 0.921 & \textbf{0.998} & 0.395 & 0.986 & 0.837 & \textbf{0.992} & 0.344 & 0.956 & 0.255 & \textbf{0.991} & \textbf{0.114} \\
\toolname 8B & \textbf{0.988} & \textbf{0.928} & \textbf{0.998} & \textbf{0.454} & \textbf{0.987} & \textbf{0.850} & \textbf{0.992} & \textbf{0.360} & \textbf{0.967} & \textbf{0.343} & \textbf{0.991} & 0.113 \\
\bottomrule
\end{tabular}
}
\end{table*}

\sstitle{Datasets}  
We evaluate on large–schema corpora:

\begin{compactitem}
\item \textit{Spider} \cite{yu2018spider} is a widely used NL2SQL benchmark with 200 databases spanning 138 domains. It provides 8,659 training examples, 1{,}024 development examples, and a test set.
\item \textit{BIRD} \cite{li2024can} has 95 real–world databases across 37 domains with 9,428 training, 1,534 development, and a hidden test set. Its databases are much larger than Spider (avg. 549K vs. 2K rows) and often include external evidence passages.  
\item \textit{Spider 2.0} \cite{lei2024spider} is a recent benchmark of 632 text-to-SQL workflow problems designed to reflect realistic enterprise scenarios with much larger and more complex schemas than earlier corpora. We use the partially released Spider 2.0-lite, selecting only databases with full public schema information to construct an evaluation dataset. This yields 233 evaluation samples spanning multiple SQL dialects (BigQuery, Snowflake, and SQLite). Databases can exceed 1,000 columns, SQL queries $\sim$100 lines, and up to 70K columns from timestamped replications. To make evaluation tractable while preserving semantic coverage, we apply a table grouping step inspired by Spider-Agent that merges tables with identical structures before column-level filtering \cite{lei2024spider}.

\end{compactitem}

\autoref{tab:schema-stats-4sets} summarizes schema statistics for all datasets, including our constructed Spider 2.0–lite evaluation set. It reports medians, 95th percentiles, and maxima for the numbers of tables and columns, showing that all corpora include databases with very large schemas, underscoring the need for scalable schema filtering that maintains high recall while suppressing irrelevant fields.

\sstitle{Baselines}  
We compare against schema-filtering families covering ranking, graph/pruning, LLM ranking, and embedding retrieval:

\begin{compactitem}
\item SchemaExP (Schema Expansion \& Pruning)~\cite{zhao2022bridging}: a modular pre-processing pipeline that expands candidate fields and then prunes the schema for tractable decoding. We use the pruning stage as a filter. 

  \item CodeS~\cite{li2024codes}: 350M and 3.5B open-source Text2SQL models extending RESDSQL~\cite{li2023resdsql}. We use their schema-filter variants as scoring-based rankers for per-column and per-table relevance. Following the default setup, CodeS selects the top-6 tables and top-10 columns on Spider and BIRD. We choose top-40 tables and top-80 columns on Spider~2.0-Lite for high recall. The ``+ chunking'' setting divides schema elements into batches to fit token limits.

  \item BGE Reranker~\cite{chen2024m3}: cross-encoder models (v2-m3 and lighter v2-miniCPM) that score query–column pairs with per-column probabilities, enabling thresholded selection.

  \item Qwen3 Reranker~\cite{qwen3embedding}: a state-of-the-art LLM reranker from the Qwen3 series, widely used in RAG systems, that assigns query–column relevance scores.

  \item Embedding-based methods: we adopt embedding retrievers (BGE-m3~\cite{chen2024m3} and Qwen3-Embedding-0.6B~\cite{qwen3embedding}) that rank columns by query-column embedding similarity.

\end{compactitem}

\sstitle{Metrics}  
We evaluate scoring methods assigning per-column relevance, including CodeS, LM rerankers, embedding models, and our approach. We report ROC AUC (the main metric for comparing methods) and PR AUC under threshold sweeps. We also report precision under high recall (close to 1.0), reflecting the need in Text2SQL to retain all relevant columns while suppressing noise. Note that exact recall of~1.0 is unnecessary, as some questions admit multiple SQL queries executing correctly beyond the label.

\subsection{Schema Filtering Performance}
\label{sec:mainresults}

\begin{figure*}[!h]
  \centering
  \setlength{\abovecaptionskip}{2pt}   %
    \setlength{\belowcaptionskip}{0pt}   %
  \includegraphics[width=0.98\linewidth]{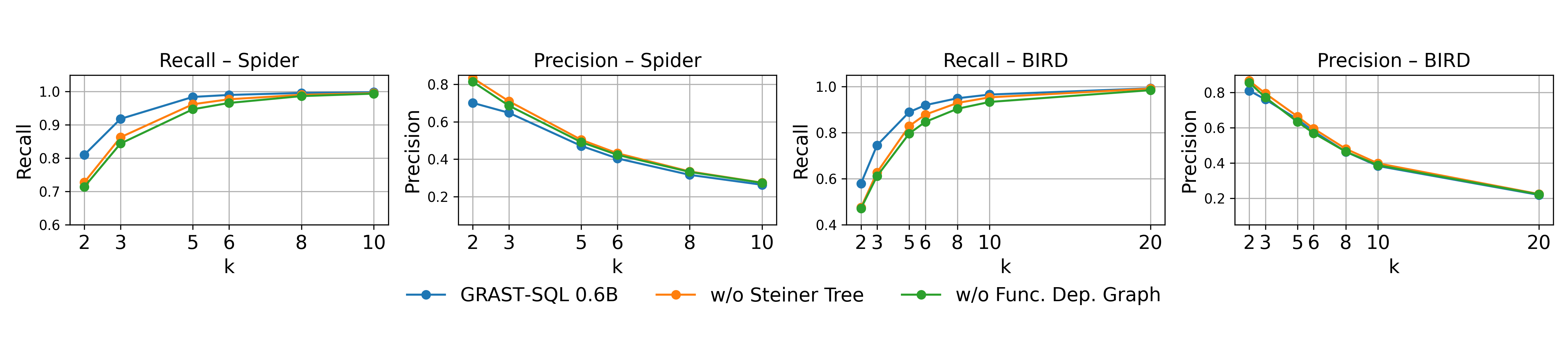}
  \vspace{-.3cm}
  \caption{Column--level recall and precision on Spider and BIRD for the full \toolname pipeline and two reduced variants.}
  \label{fig:ablation-study}
\end{figure*}

\begin{figure*}[!h]
  \centering
  \setlength{\abovecaptionskip}{2pt}   %
    \setlength{\belowcaptionskip}{0pt}   %
  \includegraphics[width=0.99\textwidth]{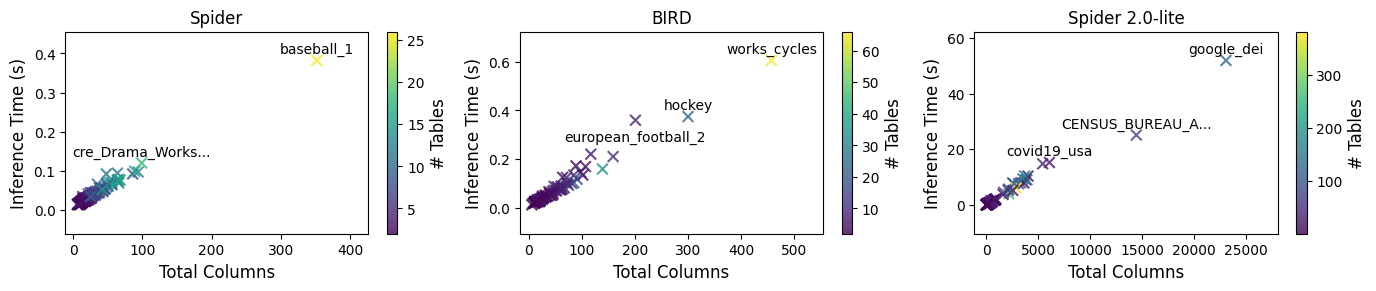}
  \caption{Inference latency vs.\ schema width (columns) on Spider, BIRD, and Spider~2.0-Lite; color encodes number of tables.}
  \label{fig:efficiency-all}
\end{figure*}

\autoref{tab:main-results} reports column-level results across Spider, BIRD, and Spider 2.0-lite. We compare schema filtering baselines (SchemaExp, BGE retrievers and rerankers, CodeS, Qwen3 rerankers) with our proposed \toolname. All scoring methods assign per-column relevance scores, so we evaluate ranking quality using ROC AUC and PR AUC under threshold sweeps. Because recall is critical in Text2SQL, omitting a necessary column typically renders the query unsolvable. We additionally report precision at a high-recall operating point, selecting a threshold that pushes recall close to \(1.0\) and then measuring the resulting precision. This setting highlights each method's ability to suppress irrelevant columns without sacrificing coverage, noting that higher recall often lowers precision.

While our method maintains consistently high recall ($0.998$ on Spider, $0.992$ on BIRD, $0.991$ on Spider 2.0-lite),  it achieves substantially higher precision than embedding retrievers (BGE-m3) and cross-encoder rerankers (BGE v2-m3, v2-miniCPM). Compared to Qwen3 rerankers and CodeS, \toolname also delivers stronger precision at this high-recall operating point, demonstrating a more favorable recall--precision balance. In addition, it attains the best ROC AUC and PR AUC across datasets (e.g., ROC AUC $0.988$ on Spider, $0.987$ on BIRD, $0.967$ on Spider 2.0-lite), confirming robustness under threshold variation. Notably, our lightweight version, \toolname 0.6B, also achieves impressive results, maintaining high recall and precision with fewer parameters, making it suitable for resource-constrained environments or when dealing with a large number of columns. These results indicate that our approach produces compact yet structurally sufficient column sets, which are essential for downstream SQL generation.

\vspace{-1.0em}
\subsection{Ablation Study}
\label{sec:ablation}

In \autoref{fig:ablation-study}, we report top--$k$ recall and precision with $k$ ranging from 2 to 20 for Spider and BIRD. We compare three configurations: (1) the full \toolname pipeline with the 0.6B model, (2) a variant without the Steiner tree expansion but retaining the graph transformer, and (3) a variant without the functional dependency graph—disabling both the graph-transformer refinement and the Steiner algorithm.  

Our ablation study reveals the distinct contributions of each component. Incorporating the Steiner expansion notably boosts low-k recall: at k{=}2, recall rises from 0.7272 to 0.8100 on Spider and from 0.4741 to 0.5790 on BIRD. As expected, this comes with a modest precision trade-off at k{=}2 (e.g., Spider precision drops from 0.8337 to 0.7010; BIRD from 0.8669 to 0.8110) due to the introduction of additional candidates. The graph modeling is critical: removing both the functional-dependency graph (and its graph-transformer refinement) and the Steiner layer (w/o~Func. Dep. Graph) yields the steepest drop at k{=}2 recall (0.7137 on Spider, 0.4705 on BIRD), underscoring the value of schema context. The LLM-reranker is not sufficient alone, as evidenced by the consistent gap between w/o Func. Dep. Graph and the full model. Overall, these results confirm the complementary strengths of the Steiner expansion, graph transformer, and the LLM in achieving strong overall performance.

\vspace{-0.8em}
\subsection{Efficiency Evaluation}
\label{sec:efficiency}

\autoref{fig:efficiency-all} shows schema-filtering latency against database size, measured in number of tables and total columns. This experiment evaluates how our model scales to increasingly large and complex schemas. In this experiment, our model was served through vLLM~\cite{kwon2023efficient}, a modern LLM inference engine with KV-cache mechanism, ensuring accelerated decoding and reduced latency. This compatibility is also an advantage of our approach: unlike CodeS and SchemaExP, which require task-specific pipelines.

On Spider, most databases are processed in well under a second, with median latency below 0.2s. Even larger cases such as \texttt{baseball\_1} (26 tables, 352 columns) complete filtering in about 0.39s.  
On BIRD, the trend is similar: nearly all databases are filtered in milliseconds, with a median latency below 0.61s. The slowest case, \texttt{works\_cycles} (66 tables, 457 columns), still completes in 0.61s.  
Finally, on the Spider~2.0–Lite public set, which includes far larger schemas, inference time grows smoothly with size. Databases such as \texttt{covid19\_usa} (21 tables, 6066 columns) and \texttt{CENSUS\_BUREAU\_ACS\_2} (78 tables, 14433 columns) complete filtering in 15.22s and 25.09s, respectively. Even the largest case, \texttt{google\_dei} (381 tables, 23,067 columns), finishes within 51.86s. These results show that our method achieves practical responsiveness for small- and medium-scale databases, while remaining scalable to extremely large schemas that would be infeasible for traditional context-based approaches.

\subsection{Sensitivity Analysis}
\label{sec:param-sensitivity}

\begin{figure}[!h]
    \vspace{-.3cm}
    \centering
    \setlength{\abovecaptionskip}{2pt}   %
    \setlength{\belowcaptionskip}{0pt}   %

    \includegraphics[width=0.48\textwidth]{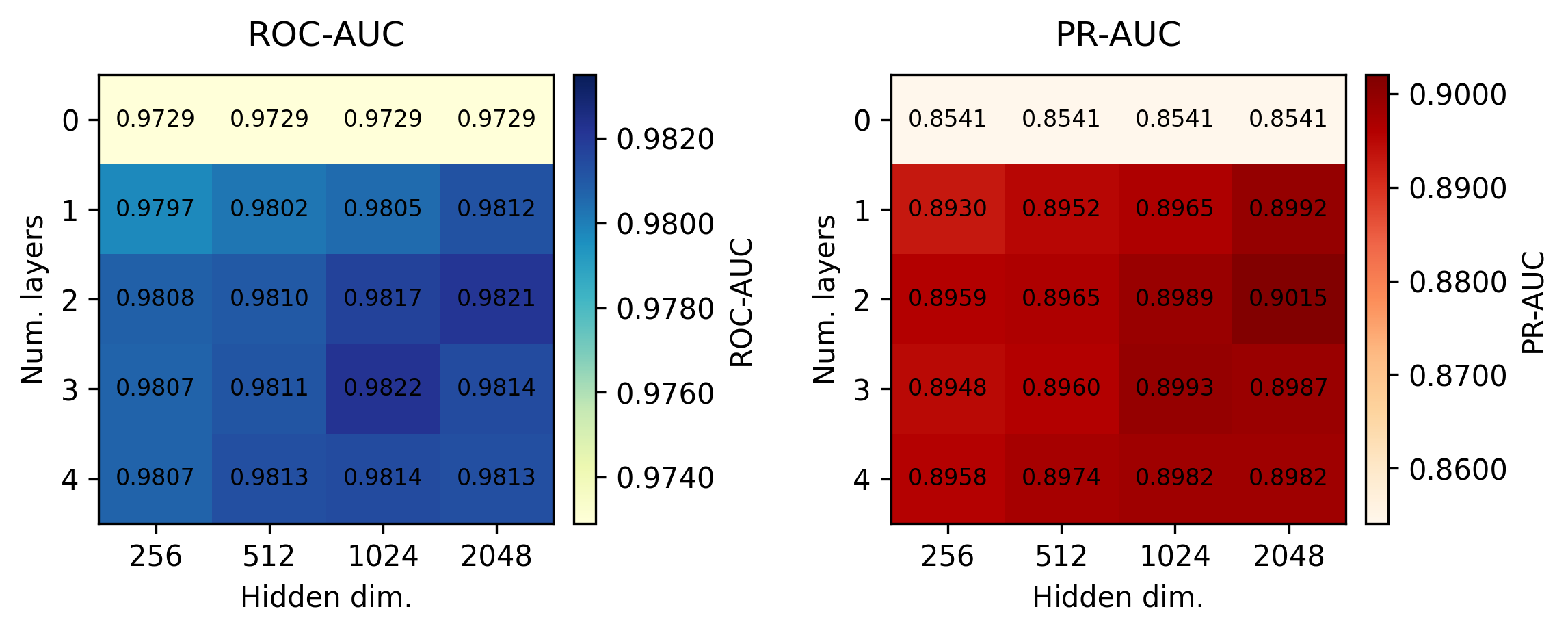}
    \vspace{-.3cm}
    \caption{ROC/PR AUC on Spider dev with \toolname 0.6B.}
    \label{fig:spider_heatmap}
    \vspace{-.1cm}
\end{figure}

\begin{figure}[!h]
    \centering
    \setlength{\abovecaptionskip}{2pt}   %
    \setlength{\belowcaptionskip}{0pt}   %
    \includegraphics[width=0.48\textwidth]{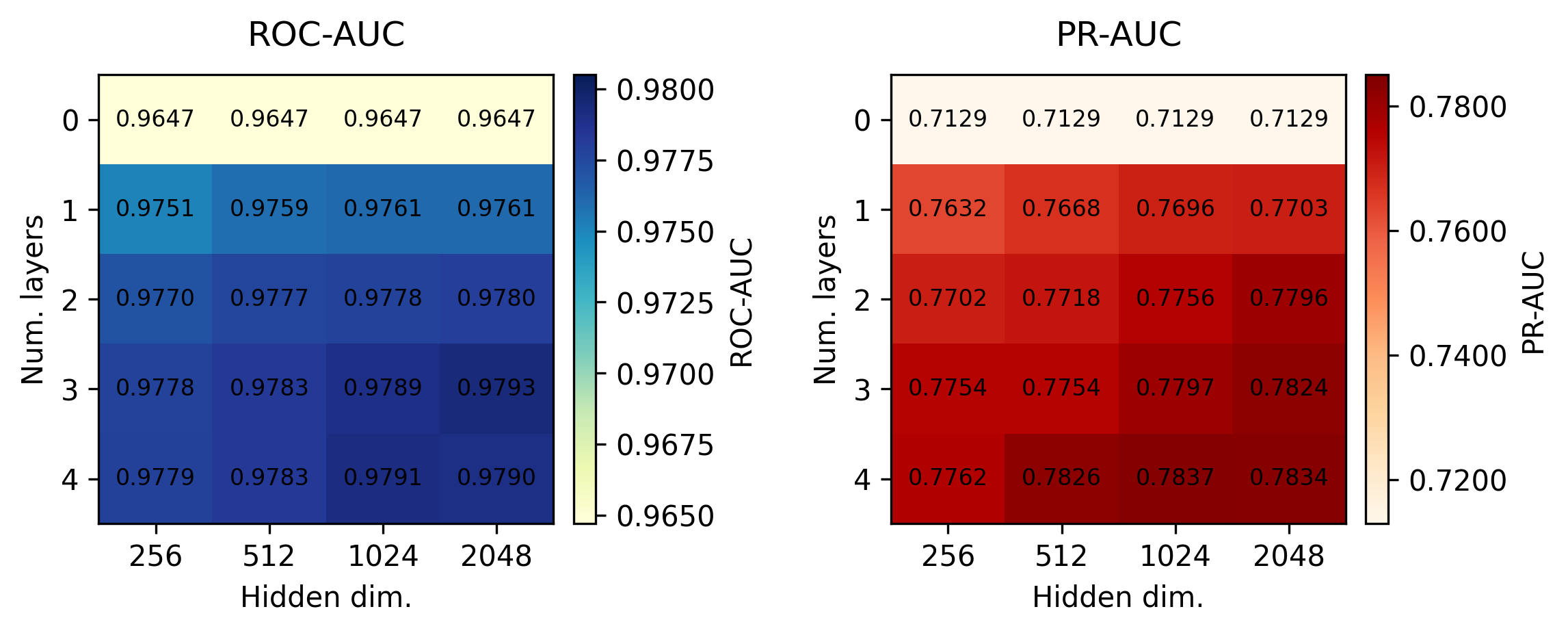}
    \vspace{-.3cm}
    \caption{ROC/PR on BIRD dev with \toolname 0.6B.}
    \label{fig:bird_heatmap}
\end{figure}

\begin{figure}[!h]
   \vspace{-.3cm}
    \centering
    \setlength{\abovecaptionskip}{2pt}   %
    \setlength{\belowcaptionskip}{0pt}   %
    \includegraphics[width=0.48\textwidth]{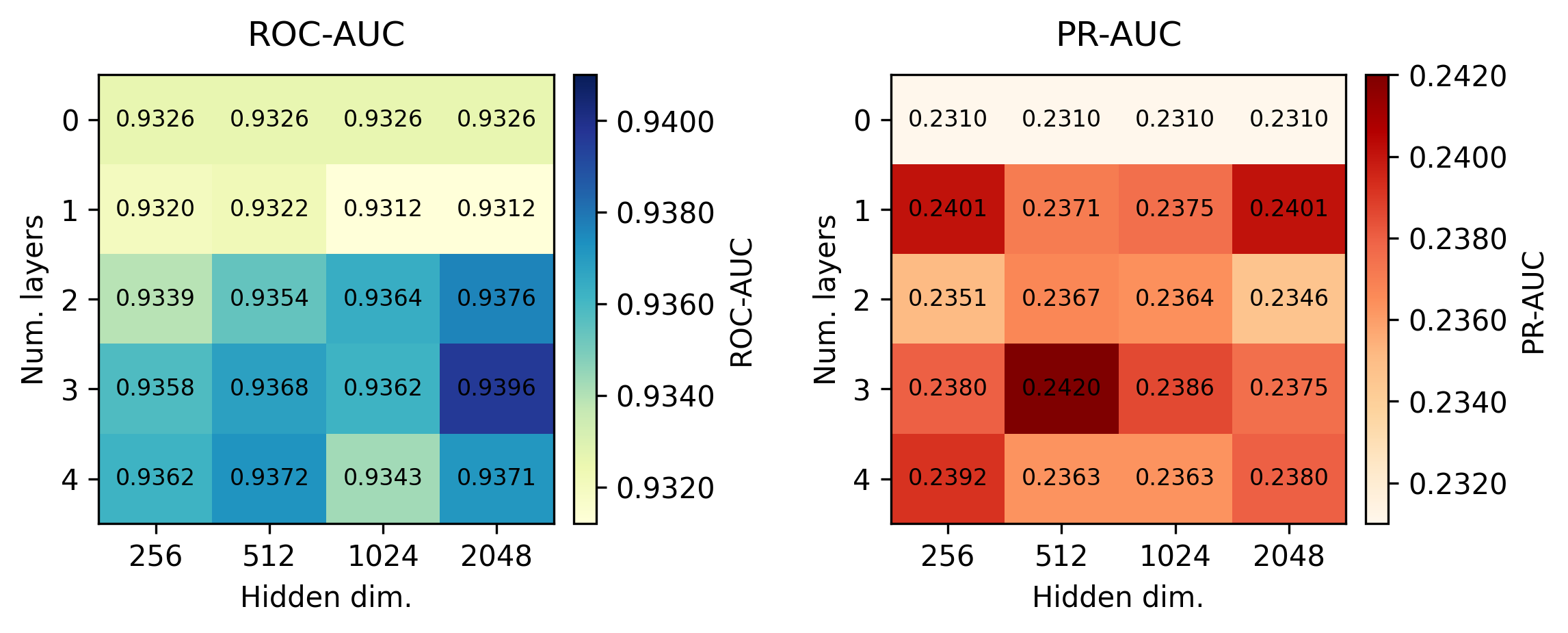}
    \vspace{-.3cm}
   \caption{ROC/PR on Spider~2.0-lite with \toolname 0.6B.}
    \label{fig:spider2_heatmap}
    \vspace{-.3cm}
\end{figure}

We conducted a grid search over two key hyperparameters: the number of Graph Transformer layers (0--4) and hidden dimension size (\{256, 512, 1024, 2048\}). \autoref{fig:spider_heatmap}, \autoref{fig:bird_heatmap} and \autoref{fig:spider2_heatmap} show ROC-AUC and PR-AUC results on the BIRD, Spider, and Spider~2.0-lite datasets.

On BIRD, performance improves with depth, stabilizing at 3--4 layers; the best ROC-AUC (0.9793) occurs with 3 layers and 2048 hidden dimension, while the top PR-AUC (0.7873) arises from a 4-layer, 1024-dimension model. On Spider, ROC-AUC peaks at 0.9822 with 3 layers and 1024 dimensions. On Spider~2.0-lite, ROC-AUC stays above 0.93 across all settings, with a maximum of 0.9396 at 3 layers and a hidden dimension of 2048, while PR-AUC peaks at 0.2420 with 3 layers. Overall, these results indicate that deeper models consistently improve ROC-AUC, though optimal PR-AUC varies by dataset. A robust configuration across benchmarks typically involves 3--4 layers with a hidden dimension of at least 1024, offering a good balance of stability and performance.

\subsection{Thresholding Analysis}
\label{sec:threshold}

\begin{figure}[!h]
  \centering
  \setlength{\abovecaptionskip}{2pt}   %
    \setlength{\belowcaptionskip}{0pt}   %
  \begin{minipage}{0.98\linewidth}
    \includegraphics[width=\linewidth]{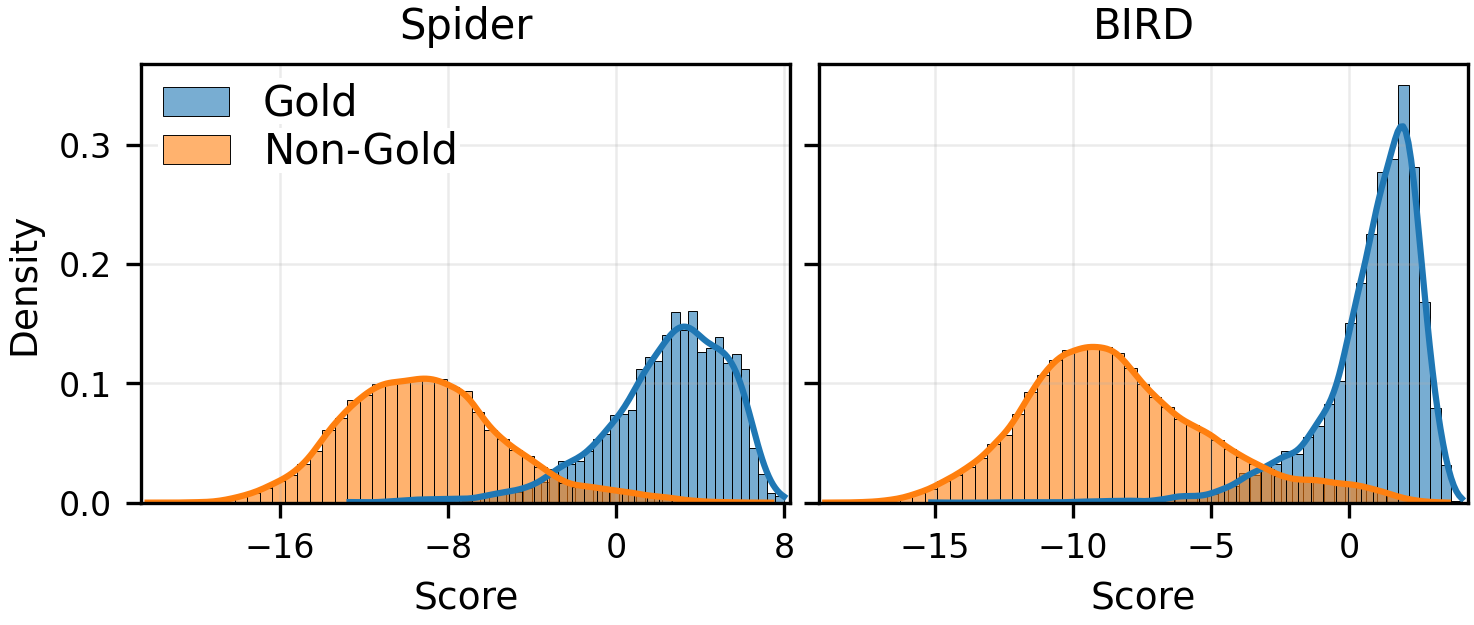}
  \end{minipage}
  \caption{Score distributions for gold vs.\ non-gold columns.}
  \label{fig:score-dists-both}
  \vspace{-.3cm}
\end{figure}

\begin{figure}[!h]
  \centering
  \setlength{\abovecaptionskip}{2pt}   %
    \setlength{\belowcaptionskip}{0pt}   %
  \begin{minipage}{0.98\linewidth}
    \includegraphics[width=\linewidth]{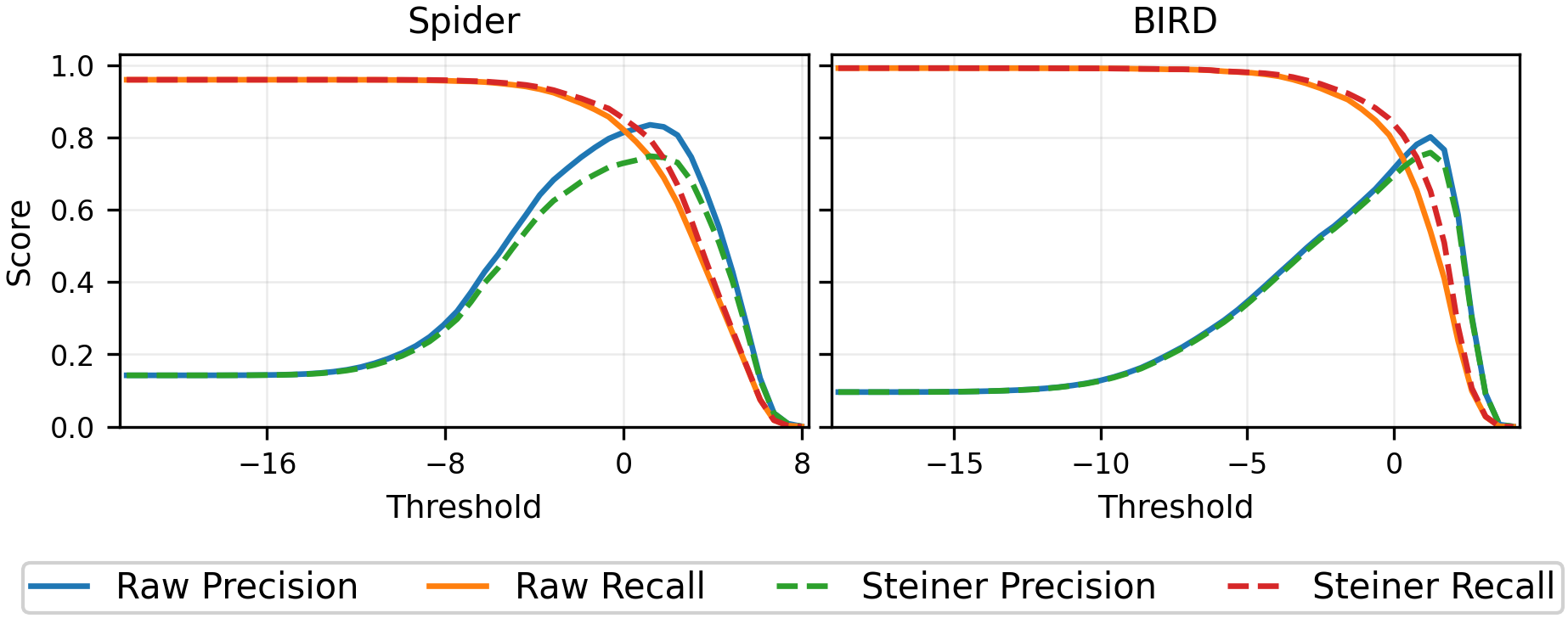}
  \end{minipage}
  \caption{Macro Precision, Recall, and F$_1$ vs.\ threshold.}
  \label{fig:macro-curves-both}
  \vspace{-0.3cm}
\end{figure}

We analyze per-column scores on Spider and BIRD by sweeping a decision threshold to compute macro Precision, Recall, and F$_1$, where columns with scores above the threshold are selected. Metrics are reported both on the raw selection and after applying the Steiner step, which restores necessary join paths. As shown in \autoref{fig:score-dists-both} and \autoref{fig:macro-curves-both}, gold and non-gold distributions are well separated with little overlap, yielding high ROC and PR AUC. Raising the threshold increases precision while lowering recall, and Steiner slightly increases recall by adding indispensable columns for join paths, which in turn stabilizes F$_1$ in the high-recall region. However, we recommend using top-$K$ or top-percentage selection instead of a fixed threshold, as the optimal cutoff can vary depending on the dataset or change after re-training.

\subsection{Text2SQL End-to-End Comparison}
\label{sec:end2end}

To evaluate the impact of \toolname on end-to-end Text2SQL performance and LLM inference cost, we compare models with and without \toolname on the Spider and BIRD development sets. As a preprocessing module for schema pruning, \toolname analyzes the question and schema to retain only the most relevant tables and columns, thereby shrinking the prompt before it is fed to the LLM. This reduction lowers token consumption, and thus API costs, without requiring additional LLM calls. For experiments, we integrate \toolname into MAC-SQL and DIN-SQL using \texttt{gpt-4.1-mini}, a cost-effective LLM sufficient to support our conclusions.

\begin{table}[!h]
\centering
\small
\vspace{-.5em}
\caption{End-to-end performance with \toolname integration. EX = execution accuracy, $P$ = prompt tokens, $R$ = response tokens, $\downarrow_P$ = prompt reduction (\%).}
\label{tab:end2end-comparison}
\vspace{-0.1cm}
\begin{tabular}{l|c|c|c|c|c|c|c|c}
\hline
\multirow{2}{*}{Model} & \multicolumn{4}{c|}{Spider Dev} & \multicolumn{4}{c}{BIRD Dev} \\
\cline{2-9}
 & EX & $P$ & $R$ & $\downarrow_P$ & EX & $P$ & $R$ & $\downarrow_P$ \\
\hline
MAC-SQL & 79.6 & 2.28M & 88.6k & {-} & 59.8 & 8.15M & 1.03M & {-} \\
+ GRAST & 80.3 & 1.59M & 84.7k & 30.2 & 59.7 & 4.07M & 939.8k & 50.0 \\
\hline
DIN-SQL & 79.0 & 9.30M & 447.3k & {-} & 58.9 & 38.7M & 1.37M & {-} \\
+ GRAST & 80.2 & 8.42M & 469.5k & 9.4 & 59.1 & 25.9M & 1.34M & 33.1 \\
\hline
\end{tabular}
\end{table}

Results in \autoref{tab:end2end-comparison} show substantial token savings with no loss of overall accuracy. On BIRD, where schemas are large, \toolname reduces prompt tokens by 50.0\% for MAC-SQL (8.1M $\rightarrow$ 4.1M) and 33.1\% for DIN-SQL (38.7M $\rightarrow$ 25.9M), with minimal accuracy change. On Spider, prompt tokens decrease by 30.2\% for MAC-SQL and 9.4\% for DIN-SQL, both with slight accuracy gains. By pruning irrelevant context, \toolname achieves up to 50\% token reduction, further cutting operational costs tied to input length while preserving or improving performance. Its lightweight design ensures compatibility with existing pipelines, providing a practical plug-in for efficient Text2SQL deployment.

\section{Conclusion}
\label{sec:conclusion}

We presented \toolname, an open-source schema filtering pipeline that ranks query-column pairs, exploits functional dependencies, and applies Steiner-tree connectivity to yield compact yet complete sub-schemas. Unlike prior work relying on independent ranking or costly long-context LLMs, our method unifies LLM scoring, graph reasoning, and connectivity guarantees. Its implementation required recovering missing keys, designing a graph-based reranker, and adapting a Steiner-tree algorithm for efficiency at schema scales of 23K+ columns. On Spider, BIRD, and Spider~2.0-lite, \toolname achieves state-of-the-art ROC AUC and PR AUC with high precision at near-perfect recall, cuts prompt tokens by up to 50\% without extra LLM calls, and runs with sub-second latency on modern inference engines.

\balance 


\begin{thebibliography}{37}


\ifx \showCODEN    \undefined \def \showCODEN     #1{\unskip}     \fi
\ifx \showDOI      \undefined \def \showDOI       #1{#1}\fi
\ifx \showISBNx    \undefined \def \showISBNx     #1{\unskip}     \fi
\ifx \showISBNxiii \undefined \def \showISBNxiii  #1{\unskip}     \fi
\ifx \showISSN     \undefined \def \showISSN      #1{\unskip}     \fi
\ifx \showLCCN     \undefined \def \showLCCN      #1{\unskip}     \fi
\ifx \shownote     \undefined \def \shownote      #1{#1}          \fi
\ifx \showarticletitle \undefined \def \showarticletitle #1{#1}   \fi
\ifx \showURL      \undefined \def \showURL       {\relax}        \fi
\providecommand\bibfield[2]{#2}
\providecommand\bibinfo[2]{#2}
\providecommand\natexlab[1]{#1}
\providecommand\showeprint[2][]{arXiv:#2}

\bibitem[Bogin et~al\mbox{.}(2019a)]%
        {bogin2019representing}
\bibfield{author}{\bibinfo{person}{Ben Bogin}, \bibinfo{person}{Jonathan
  Berant}, {and} \bibinfo{person}{Matt Gardner}.}
  \bibinfo{year}{2019}\natexlab{a}.
\newblock \showarticletitle{Representing Schema Structure with Graph Neural
  Networks for Text2SQL Parsing}. In \bibinfo{booktitle}{\emph{ACL}}.
  \bibinfo{pages}{4560--4565}.
\newblock


\bibitem[Bogin et~al\mbox{.}(2019b)]%
        {bogin2019global}
\bibfield{author}{\bibinfo{person}{Ben Bogin}, \bibinfo{person}{Matt Gardner},
  {and} \bibinfo{person}{Jonathan Berant}.} \bibinfo{year}{2019}\natexlab{b}.
\newblock \showarticletitle{Global Reasoning over Database Structures for
  Text2SQL Parsing}. In \bibinfo{booktitle}{\emph{EMNLP-IJCNLP}}.
  \bibinfo{pages}{3659--3664}.
\newblock


\bibitem[Chen et~al\mbox{.}(2024)]%
        {chen2024m3}
\bibfield{author}{\bibinfo{person}{Jianlyu Chen}, \bibinfo{person}{Shitao
  Xiao}, \bibinfo{person}{Peitian Zhang}, \bibinfo{person}{Kun Luo},
  \bibinfo{person}{Defu Lian}, {and} \bibinfo{person}{Zheng Liu}.}
  \bibinfo{year}{2024}\natexlab{}.
\newblock \showarticletitle{M3-embedding: Multi-linguality,
  multi-functionality, multi-granularity text embeddings through self-knowledge
  distillation}. In \bibinfo{booktitle}{\emph{ACL}}.
  \bibinfo{pages}{2318--2335}.
\newblock


\bibitem[Chen et~al\mbox{.}(2021)]%
        {chen2021shadowgnn}
\bibfield{author}{\bibinfo{person}{Zhi Chen}, \bibinfo{person}{Lu Chen},
  \bibinfo{person}{Yanbin Zhao}, \bibinfo{person}{Ruisheng Cao},
  \bibinfo{person}{Zihan Xu}, \bibinfo{person}{Su Zhu}, {and}
  \bibinfo{person}{Kai Yu}.} \bibinfo{year}{2021}\natexlab{}.
\newblock \showarticletitle{ShadowGNN: Graph Projection Neural Network for
  Text2SQL Parser}. In \bibinfo{booktitle}{\emph{NAACL-HLT}}.
  \bibinfo{pages}{5567--5577}.
\newblock


\bibitem[Chung et~al\mbox{.}(2025)]%
        {chung2025long}
\bibfield{author}{\bibinfo{person}{Yeounoh Chung},
  \bibinfo{person}{Gaurav~Tarlok Kakkar}, \bibinfo{person}{Yu Gan},
  \bibinfo{person}{Brenton Milne}, {and} \bibinfo{person}{Fatma Ozcan}.}
  \bibinfo{year}{2025}\natexlab{}.
\newblock \showarticletitle{Is Long Context All You Need? Leveraging LLM's
  Extended Context for {NL2SQL}}.
\newblock \bibinfo{journal}{\emph{Proc. {VLDB} Endow.}} \bibinfo{volume}{18},
  \bibinfo{number}{8} (\bibinfo{year}{2025}), \bibinfo{pages}{2735--2747}.
\newblock
\urldef\tempurl%
\url{https://www.vldb.org/pvldb/vol18/p2735-ozcan.pdf}
\showURL{%
\tempurl}


\bibitem[Deng et~al\mbox{.}(2025)]%
        {deng2025reforce}
\bibfield{author}{\bibinfo{person}{Minghang Deng}, \bibinfo{person}{Ashwin
  Ramachandran}, \bibinfo{person}{Canwen Xu}, \bibinfo{person}{Lanxiang Hu},
  \bibinfo{person}{Zhewei Yao}, \bibinfo{person}{Anupam Datta}, {and}
  \bibinfo{person}{Hao Zhang}.} \bibinfo{year}{2025}\natexlab{}.
\newblock \showarticletitle{ReFoRCE: A Text2SQL Agent with Self-Refinement,
  Consensus Enforcement, and Column Exploration}.
\newblock \bibinfo{journal}{\emph{arXiv preprint arXiv:2502.00675}}
  (\bibinfo{year}{2025}).
\newblock


\bibitem[Dong et~al\mbox{.}(2023)]%
        {dong2023c3}
\bibfield{author}{\bibinfo{person}{Xuemei Dong}, \bibinfo{person}{Chao Zhang},
  \bibinfo{person}{Yuhang Ge}, \bibinfo{person}{Yuren Mao},
  \bibinfo{person}{Yunjun Gao}, \bibinfo{person}{Jinshu Lin},
  \bibinfo{person}{Dongfang Lou}, {et~al\mbox{.}}}
  \bibinfo{year}{2023}\natexlab{}.
\newblock \showarticletitle{C3: Zero-shot text-to-sql with chatgpt}.
\newblock \bibinfo{journal}{\emph{arXiv preprint arXiv:2307.07306}}
  (\bibinfo{year}{2023}).
\newblock


\bibitem[Guo et~al\mbox{.}(2019)]%
        {guo2019towards}
\bibfield{author}{\bibinfo{person}{Jiaqi Guo}, \bibinfo{person}{Zecheng Zhan},
  \bibinfo{person}{Yan Gao}, \bibinfo{person}{Yan Xiao},
  \bibinfo{person}{Jian-Guang Lou}, \bibinfo{person}{Ting Liu}, {and}
  \bibinfo{person}{Dongmei Zhang}.} \bibinfo{year}{2019}\natexlab{}.
\newblock \showarticletitle{Towards Complex Text2SQL in Cross-Domain Database
  with Intermediate Representation}. In \bibinfo{booktitle}{\emph{ACL}}.
  \bibinfo{pages}{4524--4535}.
\newblock


\bibitem[Hwang and Richards(1992)]%
        {hwang1992steiner}
\bibfield{author}{\bibinfo{person}{Frank~K Hwang} {and} \bibinfo{person}{Dana~S
  Richards}.} \bibinfo{year}{1992}\natexlab{}.
\newblock \showarticletitle{Steiner tree problems}.
\newblock \bibinfo{journal}{\emph{Networks}} \bibinfo{volume}{22},
  \bibinfo{number}{1} (\bibinfo{year}{1992}), \bibinfo{pages}{55--89}.
\newblock


\bibitem[Kwon et~al\mbox{.}(2023)]%
        {kwon2023efficient}
\bibfield{author}{\bibinfo{person}{Woosuk Kwon}, \bibinfo{person}{Zhuohan Li},
  \bibinfo{person}{Siyuan Zhuang}, \bibinfo{person}{Ying Sheng},
  \bibinfo{person}{Lianmin Zheng}, \bibinfo{person}{Cody~Hao Yu},
  \bibinfo{person}{Joseph Gonzalez}, \bibinfo{person}{Hao Zhang}, {and}
  \bibinfo{person}{Ion Stoica}.} \bibinfo{year}{2023}\natexlab{}.
\newblock \showarticletitle{Efficient memory management for large language
  model serving with pagedattention}. In \bibinfo{booktitle}{\emph{SOSP}}.
  \bibinfo{pages}{611--626}.
\newblock


\bibitem[Lee et~al\mbox{.}(2025)]%
        {lee2025dcg}
\bibfield{author}{\bibinfo{person}{Jihyung Lee}, \bibinfo{person}{Jin-Seop
  Lee}, \bibinfo{person}{Jaehoon Lee}, \bibinfo{person}{YunSeok Choi}, {and}
  \bibinfo{person}{Jee-Hyong Lee}.} \bibinfo{year}{2025}\natexlab{}.
\newblock \showarticletitle{DCG-SQL: Enhancing In-Context Learning for Text2SQL
  with Deep Contextual Schema Link Graph}.
\newblock \bibinfo{journal}{\emph{arXiv preprint arXiv:2505.19956}}
  (\bibinfo{year}{2025}).
\newblock


\bibitem[Lei et~al\mbox{.}({[n.\,d.]})]%
        {lei2024spider}
\bibfield{author}{\bibinfo{person}{Fangyu Lei}, \bibinfo{person}{Jixuan Chen},
  \bibinfo{person}{Yuxiao Ye}, \bibinfo{person}{Ruisheng Cao},
  \bibinfo{person}{Dongchan Shin}, \bibinfo{person}{Hongjin SU},
  \bibinfo{person}{ZHAOQING SUO}, \bibinfo{person}{Hongcheng Gao},
  \bibinfo{person}{Wenjing Hu}, \bibinfo{person}{Pengcheng Yin},
  {et~al\mbox{.}}} \bibinfo{year}{[n.\,d.]}\natexlab{}.
\newblock \showarticletitle{Spider 2.0: Evaluating Language Models on
  Real-World Enterprise Text2SQL Workflows}. In
  \bibinfo{booktitle}{\emph{ICLR}}.
\newblock


\bibitem[Li et~al\mbox{.}(2025)]%
        {li2025alphasql}
\bibfield{author}{\bibinfo{person}{Boyan Li}, \bibinfo{person}{Jiayi Zhang},
  \bibinfo{person}{Ju Fan}, \bibinfo{person}{Yanwei Xu}, \bibinfo{person}{Chong
  Chen}, \bibinfo{person}{Nan Tang}, {and} \bibinfo{person}{Yuyu Luo}.}
  \bibinfo{year}{2025}\natexlab{}.
\newblock \showarticletitle{Alpha-{SQL}: Zero-Shot Text-to-{SQL} using Monte
  Carlo Tree Search}. In \bibinfo{booktitle}{\emph{ICML}}.
\newblock


\bibitem[Li et~al\mbox{.}(2023b)]%
        {li2023resdsql}
\bibfield{author}{\bibinfo{person}{Haoyang Li}, \bibinfo{person}{Jing Zhang},
  \bibinfo{person}{Cuiping Li}, {and} \bibinfo{person}{Hong Chen}.}
  \bibinfo{year}{2023}\natexlab{b}.
\newblock \showarticletitle{Resdsql: Decoupling schema linking and skeleton
  parsing for text-to-sql}. In \bibinfo{booktitle}{\emph{AAAI}},
  Vol.~\bibinfo{volume}{37}. \bibinfo{pages}{13067--13075}.
\newblock


\bibitem[Li et~al\mbox{.}(2024b)]%
        {li2024codes}
\bibfield{author}{\bibinfo{person}{Haoyang Li}, \bibinfo{person}{Jing Zhang},
  \bibinfo{person}{Hanbing Liu}, \bibinfo{person}{Ju Fan},
  \bibinfo{person}{Xiaokang Zhang}, \bibinfo{person}{Jun Zhu},
  \bibinfo{person}{Renjie Wei}, \bibinfo{person}{Hongyan Pan},
  \bibinfo{person}{Cuiping Li}, {and} \bibinfo{person}{Hong Chen}.}
  \bibinfo{year}{2024}\natexlab{b}.
\newblock \showarticletitle{Codes: Towards building open-source language models
  for text-to-sql}.
\newblock \bibinfo{journal}{\emph{PACMMOD}} \bibinfo{volume}{2},
  \bibinfo{number}{3} (\bibinfo{year}{2024}), \bibinfo{pages}{1--28}.
\newblock


\bibitem[Li et~al\mbox{.}(2024a)]%
        {li2024can}
\bibfield{author}{\bibinfo{person}{Jinyang Li}, \bibinfo{person}{Binyuan Hui},
  \bibinfo{person}{Ge Qu}, \bibinfo{person}{Jiaxi Yang},
  \bibinfo{person}{Binhua Li}, \bibinfo{person}{Bowen Li},
  \bibinfo{person}{Bailin Wang}, \bibinfo{person}{Bowen Qin},
  \bibinfo{person}{Ruiying Geng}, \bibinfo{person}{Nan Huo}, {et~al\mbox{.}}}
  \bibinfo{year}{2024}\natexlab{a}.
\newblock \showarticletitle{Can llm already serve as a database interface? a
  big bench for large-scale database grounded text-to-sqls}.
\newblock \bibinfo{journal}{\emph{NeurIPS}}  \bibinfo{volume}{36}
  (\bibinfo{year}{2024}).
\newblock


\bibitem[Li et~al\mbox{.}(2023a)]%
        {li2023starcoder}
\bibfield{author}{\bibinfo{person}{Raymond Li}, \bibinfo{person}{Loubna~Ben
  Allal}, \bibinfo{person}{Yangtian Zi}, \bibinfo{person}{Niklas Muennighoff},
  {et~al\mbox{.}}} \bibinfo{year}{2023}\natexlab{a}.
\newblock \showarticletitle{Starcoder: may the source be with you!}
\newblock \bibinfo{journal}{\emph{arXiv preprint arXiv:2305.06161}}
  (\bibinfo{year}{2023}).
\newblock


\bibitem[Lin et~al\mbox{.}(2020)]%
        {lin2020bridging}
\bibfield{author}{\bibinfo{person}{Xi~Victoria Lin}, \bibinfo{person}{Richard
  Socher}, {and} \bibinfo{person}{Caiming Xiong}.}
  \bibinfo{year}{2020}\natexlab{}.
\newblock \showarticletitle{Bridging Textual and Tabular Data for Cross-Domain
  Text2SQL Semantic Parsing}. In \bibinfo{booktitle}{\emph{EMNLP}}.
  \bibinfo{pages}{4870--4888}.
\newblock


\bibitem[Liu et~al\mbox{.}(2023)]%
        {liu2023comprehensive}
\bibfield{author}{\bibinfo{person}{Aiwei Liu}, \bibinfo{person}{Xuming Hu},
  \bibinfo{person}{Lijie Wen}, {and} \bibinfo{person}{Philip~S Yu}.}
  \bibinfo{year}{2023}\natexlab{}.
\newblock \showarticletitle{A comprehensive evaluation of ChatGPT's zero-shot
  Text2SQL capability}.
\newblock \bibinfo{journal}{\emph{arXiv preprint arXiv:2303.13547}}
  (\bibinfo{year}{2023}).
\newblock


\bibitem[Pourreza et~al\mbox{.}(2025)]%
        {pourreza2025chasesql}
\bibfield{author}{\bibinfo{person}{Mohammadreza Pourreza},
  \bibinfo{person}{Hailong Li}, \bibinfo{person}{Ruoxi Sun},
  \bibinfo{person}{Yeounoh Chung}, \bibinfo{person}{Shayan Talaei},
  \bibinfo{person}{Gaurav~Tarlok Kakkar}, \bibinfo{person}{Yu Gan},
  \bibinfo{person}{Amin Saberi}, \bibinfo{person}{Fatma Ozcan}, {and}
  \bibinfo{person}{Sercan~O Arik}.} \bibinfo{year}{2025}\natexlab{}.
\newblock \showarticletitle{{CHASE}-{SQL}: Multi-Path Reasoning and Preference
  Optimized Candidate Selection in Text-to-{SQL}}. In
  \bibinfo{booktitle}{\emph{ICLR}}.
\newblock


\bibitem[Pourreza and Rafiei(2023)]%
        {pourreza2023dinsql}
\bibfield{author}{\bibinfo{person}{Mohammadreza Pourreza} {and}
  \bibinfo{person}{Davood Rafiei}.} \bibinfo{year}{2023}\natexlab{}.
\newblock \showarticletitle{Din-sql: Decomposed in-context learning of
  text-to-sql with self-correction}.
\newblock \bibinfo{journal}{\emph{NeurIPS}}  \bibinfo{volume}{36}
  (\bibinfo{year}{2023}), \bibinfo{pages}{36339--36348}.
\newblock


\bibitem[Ren et~al\mbox{.}(2024)]%
        {ren2024purple}
\bibfield{author}{\bibinfo{person}{Tonghui Ren}, \bibinfo{person}{Yuankai Fan},
  \bibinfo{person}{Zhenying He}, \bibinfo{person}{Ren Huang},
  \bibinfo{person}{Jiaqi Dai}, \bibinfo{person}{Can Huang},
  \bibinfo{person}{Yinan Jing}, \bibinfo{person}{Kai Zhang},
  \bibinfo{person}{Yifan Yang}, {and} \bibinfo{person}{X~Sean Wang}.}
  \bibinfo{year}{2024}\natexlab{}.
\newblock \showarticletitle{Purple: Making a large language model a better sql
  writer}. In \bibinfo{booktitle}{\emph{ICDE}}. IEEE, \bibinfo{pages}{15--28}.
\newblock


\bibitem[Rokis and Kirikova(2022)]%
        {rokis2022challenges}
\bibfield{author}{\bibinfo{person}{Karlis Rokis} {and} \bibinfo{person}{Marite
  Kirikova}.} \bibinfo{year}{2022}\natexlab{}.
\newblock \showarticletitle{Challenges of low-code/no-code software
  development: A literature review}. In \bibinfo{booktitle}{\emph{BIR}}.
  \bibinfo{pages}{3--17}.
\newblock


\bibitem[Roziere et~al\mbox{.}(2023)]%
        {roziere2023code}
\bibfield{author}{\bibinfo{person}{Baptiste Roziere}, \bibinfo{person}{Jonas
  Gehring}, \bibinfo{person}{Fabian Gloeckle}, \bibinfo{person}{Sten Sootla},
  \bibinfo{person}{Itai Gat}, {et~al\mbox{.}}} \bibinfo{year}{2023}\natexlab{}.
\newblock \showarticletitle{Code llama: Open foundation models for code}.
\newblock \bibinfo{journal}{\emph{arXiv preprint arXiv:2308.12950}}
  (\bibinfo{year}{2023}).
\newblock


\bibitem[Scholak et~al\mbox{.}(2021)]%
        {scholak2021picard}
\bibfield{author}{\bibinfo{person}{Torsten Scholak}, \bibinfo{person}{Nathan
  Schucher}, {and} \bibinfo{person}{Dzmitry Bahdanau}.}
  \bibinfo{year}{2021}\natexlab{}.
\newblock \showarticletitle{PICARD: Parsing Incrementally for Constrained
  Auto-Regressive Decoding from Language Models}. In
  \bibinfo{booktitle}{\emph{EMNLP}}. \bibinfo{pages}{9895--9901}.
\newblock


\bibitem[Sun et~al\mbox{.}(2023)]%
        {sun2023sqlpalm}
\bibfield{author}{\bibinfo{person}{Ruoxi Sun}, \bibinfo{person}{Sercan~{\"O}
  Arik}, \bibinfo{person}{Alex Muzio}, \bibinfo{person}{Lesly Miculicich},
  \bibinfo{person}{Satya Gundabathula}, \bibinfo{person}{Pengcheng Yin},
  \bibinfo{person}{Hanjun Dai}, \bibinfo{person}{Hootan Nakhost},
  \bibinfo{person}{Rajarishi Sinha}, \bibinfo{person}{Zifeng Wang},
  {et~al\mbox{.}}} \bibinfo{year}{2023}\natexlab{}.
\newblock \showarticletitle{Sql-palm: Improved large language model adaptation
  for text-to-sql (extended)}.
\newblock \bibinfo{journal}{\emph{arXiv preprint arXiv:2306.00739}}
  (\bibinfo{year}{2023}).
\newblock


\bibitem[Talaei et~al\mbox{.}(2024)]%
        {talaei2024chess}
\bibfield{author}{\bibinfo{person}{Shayan Talaei},
  \bibinfo{person}{Mohammadreza Pourreza}, \bibinfo{person}{Yu-Chen Chang},
  \bibinfo{person}{Azalia Mirhoseini}, {and} \bibinfo{person}{Amin Saberi}.}
  \bibinfo{year}{2024}\natexlab{}.
\newblock \showarticletitle{Chess: Contextual harnessing for efficient sql
  synthesis}.
\newblock \bibinfo{journal}{\emph{arXiv preprint arXiv:2405.16755}}
  (\bibinfo{year}{2024}).
\newblock


\bibitem[Wang et~al\mbox{.}(2025)]%
        {wang2025mac}
\bibfield{author}{\bibinfo{person}{Bing Wang}, \bibinfo{person}{Changyu Ren},
  \bibinfo{person}{Jian Yang}, \bibinfo{person}{Xinnian Liang},
  \bibinfo{person}{Jiaqi Bai}, \bibinfo{person}{Linzheng Chai},
  \bibinfo{person}{Zhao Yan}, \bibinfo{person}{Qian-Wen Zhang},
  \bibinfo{person}{Di Yin}, \bibinfo{person}{Xing Sun}, {et~al\mbox{.}}}
  \bibinfo{year}{2025}\natexlab{}.
\newblock \showarticletitle{MAC-SQL: A Multi-Agent Collaborative Framework for
  Text2SQL}. In \bibinfo{booktitle}{\emph{COLING}}. \bibinfo{pages}{540--557}.
\newblock


\bibitem[Wang et~al\mbox{.}(2020)]%
        {wang2020rat}
\bibfield{author}{\bibinfo{person}{Bailin Wang}, \bibinfo{person}{Richard
  Shin}, \bibinfo{person}{Xiaodong Liu}, \bibinfo{person}{Oleksandr Polozov},
  {and} \bibinfo{person}{Matthew Richardson}.} \bibinfo{year}{2020}\natexlab{}.
\newblock \showarticletitle{RAT-SQL: Relation-Aware Schema Encoding and Linking
  for Text2SQL Parsers}. In \bibinfo{booktitle}{\emph{ACL}}.
  \bibinfo{pages}{7567--7578}.
\newblock


\bibitem[Xu et~al\mbox{.}(2017)]%
        {xu2017sqlnet}
\bibfield{author}{\bibinfo{person}{Xiaojun Xu}, \bibinfo{person}{Chang Liu},
  {and} \bibinfo{person}{Dawn Song}.} \bibinfo{year}{2017}\natexlab{}.
\newblock \showarticletitle{Sqlnet: Generating structured queries from natural
  language without reinforcement learning}.
\newblock \bibinfo{journal}{\emph{arXiv preprint arXiv:1711.04436}}
  (\bibinfo{year}{2017}).
\newblock


\bibitem[Yu et~al\mbox{.}(2018a)]%
        {yu2018typesql}
\bibfield{author}{\bibinfo{person}{Tao Yu}, \bibinfo{person}{Zifan Li},
  \bibinfo{person}{Zilin Zhang}, \bibinfo{person}{Rui Zhang}, {and}
  \bibinfo{person}{Dragomir Radev}.} \bibinfo{year}{2018}\natexlab{a}.
\newblock \showarticletitle{{T}ype{SQL}: Knowledge-Based Type-Aware Neural
  Text-to-{SQL} Generation}. In \bibinfo{booktitle}{\emph{NAACL-HLT}}.
  \bibinfo{pages}{588--594}.
\newblock


\bibitem[Yu et~al\mbox{.}(2018b)]%
        {yu2018syntaxsqlnet}
\bibfield{author}{\bibinfo{person}{Tao Yu}, \bibinfo{person}{Michihiro
  Yasunaga}, \bibinfo{person}{Kai Yang}, \bibinfo{person}{Rui Zhang},
  \bibinfo{person}{Dongxu Wang}, \bibinfo{person}{Zifan Li}, {and}
  \bibinfo{person}{Dragomir Radev}.} \bibinfo{year}{2018}\natexlab{b}.
\newblock \showarticletitle{SyntaxSQLNet: Syntax Tree Networks for Complex and
  Cross-Domain Text2SQL Task}. In \bibinfo{booktitle}{\emph{EMNLP}}.
  \bibinfo{pages}{1653--1663}.
\newblock


\bibitem[Yu et~al\mbox{.}(2018c)]%
        {yu2018spider}
\bibfield{author}{\bibinfo{person}{Tao Yu}, \bibinfo{person}{Rui Zhang},
  \bibinfo{person}{Kai Yang}, \bibinfo{person}{Michihiro Yasunaga},
  \bibinfo{person}{Dongxu Wang}, \bibinfo{person}{Zifan Li},
  \bibinfo{person}{James Ma}, \bibinfo{person}{Irene Li},
  \bibinfo{person}{Qingning Yao}, \bibinfo{person}{Shanelle Roman},
  {et~al\mbox{.}}} \bibinfo{year}{2018}\natexlab{c}.
\newblock \showarticletitle{Spider: A Large-Scale Human-Labeled Dataset for
  Complex and Cross-Domain Semantic Parsing and Text2SQL Task}. In
  \bibinfo{booktitle}{\emph{EMNLP}}. \bibinfo{pages}{3911--3921}.
\newblock


\bibitem[Yun et~al\mbox{.}(2019)]%
        {yun2019graph}
\bibfield{author}{\bibinfo{person}{Seongjun Yun}, \bibinfo{person}{Minbyul
  Jeong}, \bibinfo{person}{Raehyun Kim}, \bibinfo{person}{Jaewoo Kang}, {and}
  \bibinfo{person}{Hyunwoo~J Kim}.} \bibinfo{year}{2019}\natexlab{}.
\newblock \showarticletitle{Graph transformer networks}.
\newblock \bibinfo{journal}{\emph{NeurIPS}}  \bibinfo{volume}{32}
  (\bibinfo{year}{2019}).
\newblock


\bibitem[Zhang et~al\mbox{.}(2025)]%
        {qwen3embedding}
\bibfield{author}{\bibinfo{person}{Yanzhao Zhang}, \bibinfo{person}{Mingxin
  Li}, \bibinfo{person}{Dingkun Long}, \bibinfo{person}{Xin Zhang},
  \bibinfo{person}{Huan Lin}, \bibinfo{person}{Baosong Yang},
  \bibinfo{person}{Pengjun Xie}, \bibinfo{person}{An Yang},
  \bibinfo{person}{Dayiheng Liu}, \bibinfo{person}{Junyang Lin},
  \bibinfo{person}{Fei Huang}, {and} \bibinfo{person}{Jingren Zhou}.}
  \bibinfo{year}{2025}\natexlab{}.
\newblock \showarticletitle{Qwen3 Embedding: Advancing Text Embedding and
  Reranking Through Foundation Models}.
\newblock \bibinfo{journal}{\emph{arXiv preprint arXiv:2506.05176}}
  (\bibinfo{year}{2025}).
\newblock


\bibitem[Zhao et~al\mbox{.}(2022)]%
        {zhao2022bridging}
\bibfield{author}{\bibinfo{person}{Chen Zhao}, \bibinfo{person}{Yu Su},
  \bibinfo{person}{Adam Pauls}, {and} \bibinfo{person}{Emmanouil~Antonios
  Platanios}.} \bibinfo{year}{2022}\natexlab{}.
\newblock \showarticletitle{Bridging the generalization gap in text-to-SQL
  parsing with schema expansion}. In \bibinfo{booktitle}{\emph{ACL}}.
  \bibinfo{pages}{5568--5578}.
\newblock


\bibitem[Zhong et~al\mbox{.}(2017)]%
        {zhong2017seq2sql}
\bibfield{author}{\bibinfo{person}{Victor Zhong}, \bibinfo{person}{Caiming
  Xiong}, {and} \bibinfo{person}{Richard Socher}.}
  \bibinfo{year}{2017}\natexlab{}.
\newblock \showarticletitle{Seq2sql: Generating structured queries from natural
  language using reinforcement learning}.
\newblock \bibinfo{journal}{\emph{arXiv preprint arXiv:1709.00103}}
  (\bibinfo{year}{2017}).
\newblock


\end{thebibliography}


\end{document}